\documentclass[aps,prl,twocolumn,showpacs,preprintnumbers,amsmath,amssymb,longbibliography,superscriptaddress]{revtex4-2}

\usepackage[makeroom]{cancel}

\usepackage{amsmath}    % need for subequations
\usepackage{amssymb,bm}
\usepackage{graphicx}   % need for figures
\usepackage{verbatim}   % useful for program listings
\usepackage{color}      % use if color is used in text
\usepackage{hyperref}   % use for hypertext links, including those to external documents and URLs
\usepackage[normalem]{ulem}
\usepackage{natbib}
\usepackage{enumitem}
\usepackage{empheq}

\definecolor{darkgreen}{rgb}{0,0.5,0}
\definecolor{purple}{rgb}{0.35,0,0.35}
\definecolor{orange}{rgb}{1,0.5,0}
\definecolor{darkred}{rgb}{.7,0,0}
\definecolor{darkblue}{rgb}{0,0,.3}
\definecolor{grey}{rgb}{.6,.6,.6}
\definecolor{dimgreen}{rgb}{0.2,0.6,0.1}

\definecolor{darkgreen}{rgb}{0,0.5,0}

\begin{document}

\newcommand{\jav}[1]{{\color{red}#1}}

%\title{Fractionalized Shiba state in the attractive Hubbard model}
\title{Spectral properties of fractionalized Shiba states}
 \author{C\u at\u alin Pa\c scu Moca}
 \affiliation{HUN-REN—BME Quantum Dynamics and Correlations Research Group,
 Budapest University of Technology and Economics, M\H uegyetem rkp. 3., H-1111 Budapest, Hungary}
 \affiliation{Department of Physics, University of Oradea, 410087, Oradea, Romania}
 \author{Csan\'ad Hajd\'u}
 \affiliation{Department of Theoretical Physics, Institute of Physics,
 Budapest University of Technology and Economics, M\H uegyetem rkp. 3., H-1111 Budapest, Hungary}
 \author{Bal\' azs D\'ora}
 \affiliation{Department of Theoretical Physics, Institute of Physics,
 Budapest University of Technology and Economics, M\H uegyetem rkp. 3., H-1111 Budapest, Hungary}
 \author{Gergely Zar\'and}
 \affiliation{Department of Theoretical Physics, Institute of Physics,
 Budapest University of Technology and Economics, M\H uegyetem rkp. 3., H-1111 Budapest, Hungary}
 \affiliation{HUN-REN—BME Quantum Dynamics and Correlations Research Group,
 Budapest University of Technology and Economics, M\H uegyetem rkp. 3., H-1111 Budapest, Hungary}

\date{\today}% It is always \today, today,

\begin{abstract}
A magnetic impurity in a BCS superconductor induces the formation of a Shiba state and drives a local quantum phase transition. 
We generalize this concept to a one-dimensional superconductor with fractionalized excitations, where the dominant instability is superconducting.
 In this framework, conduction electrons fractionalize into gapless charge and gapped spin excitations. We show that magnetic impurity interacts 
exclusively with the spin degrees of freedom and induces a quantum phase transition. Furthermore, charge excitations influence dynamical observables, 
giving rise to the phenomenon we term the \emph{fractionalized Shiba state}.  
At zero temperature, the tunneling spectrum exhibits universal power-law scaling with an exponent of $-1/2$ at half filling, 
stemming from the gapless charge modes that form a standard Luttinger liquid. Extending this analysis to finite temperatures 
reveals that the spectral features retain universal behavior at the critical point. 
\end{abstract}

\maketitle
\paragraph{Introduction.}

Shiba-Yu-Rusinov states~\cite{luh1965,shiba1968,rusinov1969}, also known as Shiba or Shiba-Andreev states, have attracted increasing attention in recent years. These subgap states arise in superconductors as a result of magnetic impurities, which trap antiferromagnetically aligned quasiparticles~\cite{ruby2015tunneling,villas2021tunneling}. Coupled Shiba states hold great potential as a quantum-computational platform~\cite{Chtchelkatchev2003,sau2012realizing, mishra2021yu,scherubl2020}, while their chains and higher-dimensional configurations offer a pathway to realizing interacting topological phases~\cite{roushan2009topological,seo2010transmission,xiu2011manipulating}.

Three-dimensional superconductors exist in a state of spontaneous symmetry-breaking with a 
well-defined superconducting phase.  In one dimension, however,  quantum fluctuations destroy 
long-range order. Charge and spin sectors separate, 
and have completely different characteristic properties. Electrons still form 
Cooper pairs, as signaled by a finite spin excitation gap, $\Delta_s$, but the charge sector remains 
gapless and gives rise to Luttinger liquid-like physics~\cite{essler2005,gogolin,giamarchi,cazalillarmp}. 
Similar to bulk excitations, Shiba states are therefore expected 
to \emph{fractionalize} in one dimension~\cite{pasnoori2022};
an electron tunneling into the Shiba state necessarily decomposes  into 
spin and charge, and while its spin can form a bound mid-(spin)gap state, its charge 
is dissolved in a (charge) Luttinger Liquid. 

We demonstrate that the magnetic impurity interacts exclusively with the spin degrees of freedom and  leaves the charge sector unaffected.
However, since the original electrons fractionalize into spin and charge modes, both sectors influence the properties of the system. 
As a result, we find that the usual Shiba resonances 
turn into \emph{continuous midgap singularities} in the tunneling spectrum~\cite{costi2000} which exhibits a universal 
scaling. 
\begin{figure}[t]
    \centering
    \includegraphics[width=0.9\columnwidth]{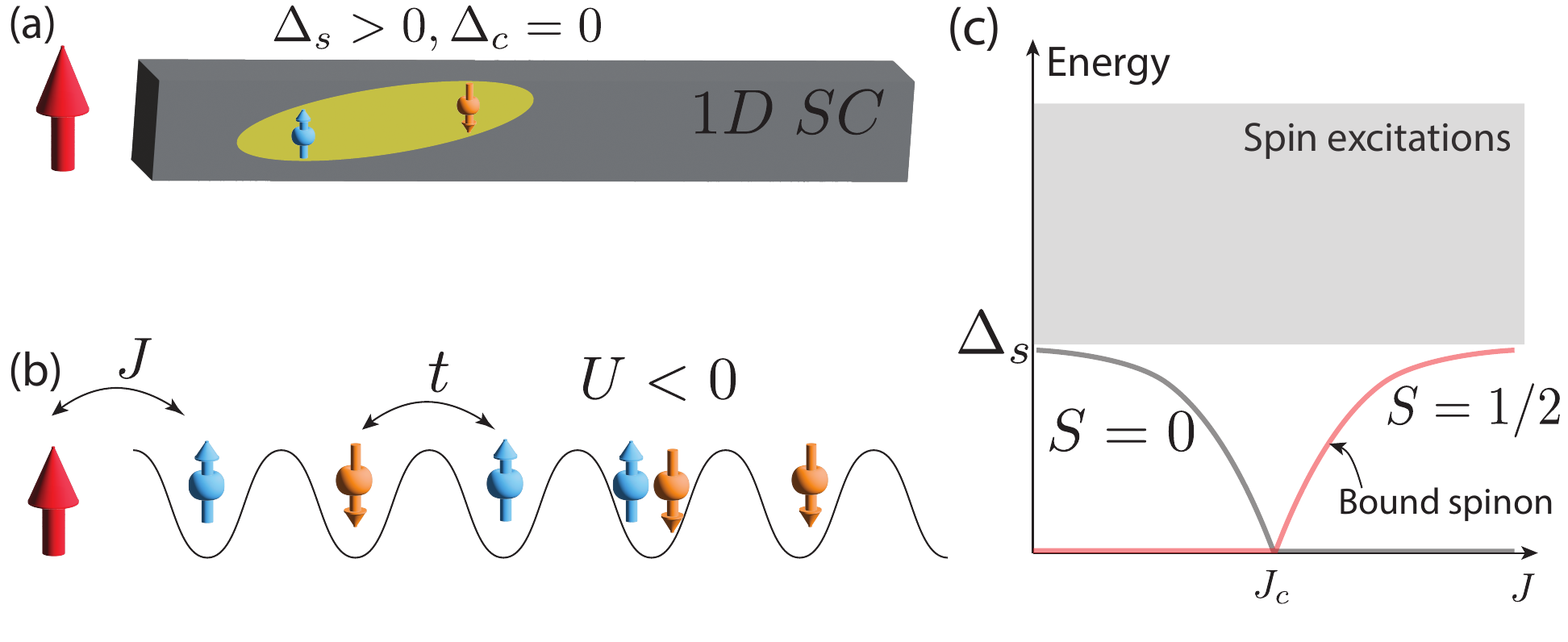} 
    \caption{(a) Typical set-up: a localized spin (large red arrow) is coupled to a one dimensional chain with dominant superconducting correlations. 
    (b) Sketch of the Hubbard model used to simulate the one dimensional superconductor. $J$ denotes the coupling between the impurity spin and the first site along the chain, 
$t$ is the hopping along the chain and $U<0$ is the attractive, on-site Coulomb interaction. 
     (c) The phase diagram of the model displays a QPT occurring at a critical exchange coupling
     $J_c$. When $J>J_c$, the ground state is a many-body singlet with fully screened localized spin, while in the opposite $J<J_c$ limit, the ground state is a
 doublet. It is characterized by an asymptotically free impurity spin,  decoupled from the chain, while the excited midgap state corresponds to a bound $S=1/2$ spinon.}
    \label{fig:sketch}
\end{figure}
The energy of the mid-gap state depends on the strength of the exchange interaction between the magnetic impurity and the superconductor, and – similar to the ordinary Shiba transition~\cite{luh1965,shiba1968,rusinov1969,vonOppen2021} – this state becomes the ground state in the spin sector 
beyond a critical coupling (see Fig.~\ref{fig:sketch}(c)). 
At this point a spin and charge parity changing transition takes place. Although this transition is of first order in nature, nevertheless, the charge sector being critical, spectral properties display scaling behaviour characteristic of second order quantum phase transitions (QPTs). 

To capture the spectral properties of the fractionalized Shiba state we follow two routes: 
we use   bosonization toolkit to  construct an  effective model to capture universal properties 
of the QPT and its finite temperature properties. 
This is supplemented by Density Matrix Renormalization Group (DMRG) 
and Matrix Product States (MPS) computations performed in the framework
of attractive Hubbard model (see Fig.~\ref{fig:sketch}(b)).

% The one-dimensional Hubbard model with attractive interactions offers a fertile ground for exploring unconventional superconducting phenomena. Superconducting correlations emerge due to the formation of Cooper pairs, even in the absence of long-range order~\cite{essler2005,gogolin,giamarchi}. Additionally, in this model, the conduction electrons fractionalize into gapless charge and gapped spin excitations, making the emergence of Shiba states in such an environment an intriguing question. 

% Here we address this problem and explore the intricate interplay between superconductivity and magnetic impurities  within the context of the 
% one-dimensional attractive Hubbard model~\cite{pasnoori2022}. 

% \jav{This represents a model system of a superconductor, displaying spin-charge separation with gapless charge and
% gapped spin modes.
% We find} that the phase diagram of the model reveals an interplay 
% between superconducting correlations and the Kondo screening effect induced by the impurity spin, leading to a QPT. 
% This transition marks a change in the ground state from a regime where the impurity spin is fully screened, characteristic of a Kondo singlet state, 
% to a phase where  superconducting correlations dominate, leaving the impurity spin largely decoupled from the electronic chain.

%
\paragraph{Attractive Hubbard model.}
To model a one dimensional superconductor, we consider the Hubbard model in a semi-infinite chain  with open boundary conditions, given by
\begin{gather}
H_{\rm Hubb}=-t\sum_{i,\sigma}\big ( c^+_{i,\sigma} c_{i+1,\sigma}+c^+_{i+1,\sigma} c_{i,\sigma}\big )+\nonumber\\
+U\sum_i \left(n_{i,\uparrow}-\frac 12\right)\left(n_{i,\downarrow}-\frac 12\right),
\end{gather}
where the first term describes hopping in one dimension with $i\geq 0$ while the second stands for the Hubbard interaction.
In one spatial dimension, the model is exactly solvable and spin-charge separation occurs. 
For $U>0$, the charge sector acquires a gap and realizes a Mott insulator at half filling only, while the spin sector remains a gapless Luttinger liquid.
Away from half filling, both sectors are gapless~\cite{giamarchi,nersesyan}.
For $U<0$, on the other hand, the spin backscattering is always a relevant perturbation~\cite{shankar} and the spin sector is gapped for all fillings, while the charge sector is gapless 
and forms a Luttinger liquid.
For the attractive case, the superconducting correlation function decays the slowest~\cite{giamarchi} 
and constitute the dominant instability, though no long range superconducting order is formed due to quantum fluctuations in one dimension.
In this and only in this respect, the model still can be considered as a 1D superconductor~\cite{gogolin}.

\paragraph{Open boundary bosonization.}
 We take the continuum limit~\cite{shankar}  as $c^+_{i,\sigma}\sim \sqrt{\alpha} \Psi^+_\sigma(x)$ with $i\sim x/\alpha$ with $\alpha$ the remnant of the lattice constant
and introduce  bosonic fields~\cite{giamarchi,nersesyan,shankar,fabrizio1995} from open boundary bosonization as
\begin{gather}
\Psi^+_\sigma(x)=\sum_{r=\pm}\frac{\kappa^+_{r,\sigma}}{\sqrt{2\pi\alpha}} \exp\left(irk_Fx+ir\phi_{\sigma}(x)-i\Theta_\sigma(x)\right),
\end{gather}
and $\kappa_{\pm,\sigma}$ represents the Klein factors for the right (+) and left ($-$) movings fields with spin $\sigma$.
The charge and spin fields are defined by
\begin{gather}
\phi_{c,s}(x)=\frac{\phi_\uparrow(x)\pm \phi_\downarrow (x)}{\sqrt 2},
\Theta_{c,s}(x)=\frac{\Theta_\uparrow(x)\pm \Theta_\downarrow (x)}{\sqrt 2}.
\end{gather}
Based on these, the spin and charge sectors of the Hubbard model are described by collective bosonic excitation~\cite{shankar,mocanu} as
\begin{subequations}
\begin{gather}  
H_c=\int_0^L \frac{dx}{2\pi} v_c\left[K_c(\partial_x\Theta_c(x))^2+\frac{1}{K_c} (\partial_x\phi_c(x))^2\right],\\
\label{eq:H_c}
H_s=\int_0^L \frac{dx}{2\pi} v_s\left[K_s(\partial_x\Theta_s(x))^2+\frac{1}{K_s} (\partial_x\phi_s(x))^2\right]+\nonumber\\
+\frac{2U\alpha}{(2\pi\alpha)^2}\cos\left(\sqrt{8}\phi_s(x)\right).
\end{gather}
\label{bosonhubbard}
\end{subequations}
Here $v_{c,s}$ are the charge and spin velocities, respectively, while $K_{c,s}$ is the Luttinger liquid parameter in the respective sectors. The umklapp term is dropped in 
 the charge sector due to its irrelevance
for attractive interactions. We emphasize that Eq. \eqref{bosonhubbard} applies not only to the Hubbard model in Fig. \ref{fig:sketch}(b), but describes the universal low energy physics of a broad class
of one dimensional systems, where attractive interactions gap out the spin sector and leave the charge degrees of freedom gapless, as depicted in Fig. \ref{fig:sketch}(a).

At half filling and $U<0$, $K_c=1$ is pinned due to charge $SU(2)$ symmetry, similarly to $K_s=1$ for the $U>0$ case due to spin $SU(2)$ symmetry~\cite{essler2005}.
However, upon doping the system, $K_c$ deviates from 1 for the attractive case. For the spin sector, the cosine term is a relevant perturbation and opens up a spin
gap $\Delta_s$ in the spectrum of spin excitations.
The spin gap associated with the superconducting order parameters in the perturbative regime, where $|U|/t \ll 1$, can be approximated as

\begin{equation}
\Delta_s \approx \sqrt{u} \exp\left( -\frac{2}{\pi u} \right),
\label{eq:Delta_s}
\end{equation}
with $u = U/4t$. Additionally, at half filling, the holon velocity remains unrenormalized at $v_c \approx 2t \big|_{u \to 0}$~\cite{essler2005}, whereas only the spinon velocity is renormalized. 
%Consequently, the Luttinger parameters are \(K_c = 1\) and \(K_s < 1\).

\paragraph{Kondo coupling.}
We supplement the Hubbard model with a Kondo impurity, localized to one end of the chain described by
\begin{gather}
H_J=J\, {\bf S} \cdot {\bf s}_0,
\end{gather}
where ${\bf s}_i=(c^+_{i,\uparrow}, c^+_{i,\downarrow}) {\bm \sigma} (c_{i,\uparrow}, c_{i,\downarrow})^T$ is the conduction electron spin operator
at site $i$, $\bm\sigma$ is the vector of Pauli matrices
and  $\bf S$ denotes the localized impurity spin-$1/2$.
The full Hamiltonian is given by 
\begin{equation}
    H = H_{\rm Hubb}+H_J   
    \label{eq:H}
\end{equation}
 with $U<0$ and $J>0$. 
In the limit $U\to 0$, the model is exactly solvable by Bethe Ansatz~\cite{andrei1983}. For antiferromagnetic coupling considered here, the localized spin get screened below the characteristic Kondo temperature,  $T_K$, 
\begin{equation}
    T_K \approx f \sqrt{j}\exp\left(-\frac{1}{j}\right),
    \label{eq:T_K}
\end{equation}
where $j=\rho_0 J$ is the dimensionless exchange interaction, with $\rho_0= \pi/2t$ being the density of states at the Fermi level, and $f$ an overall prefactor or the order unity.

\paragraph{Phase diagram and the effective spin model.} 
\begin{figure}[t]
    \centering
    \includegraphics[width=0.85\columnwidth]{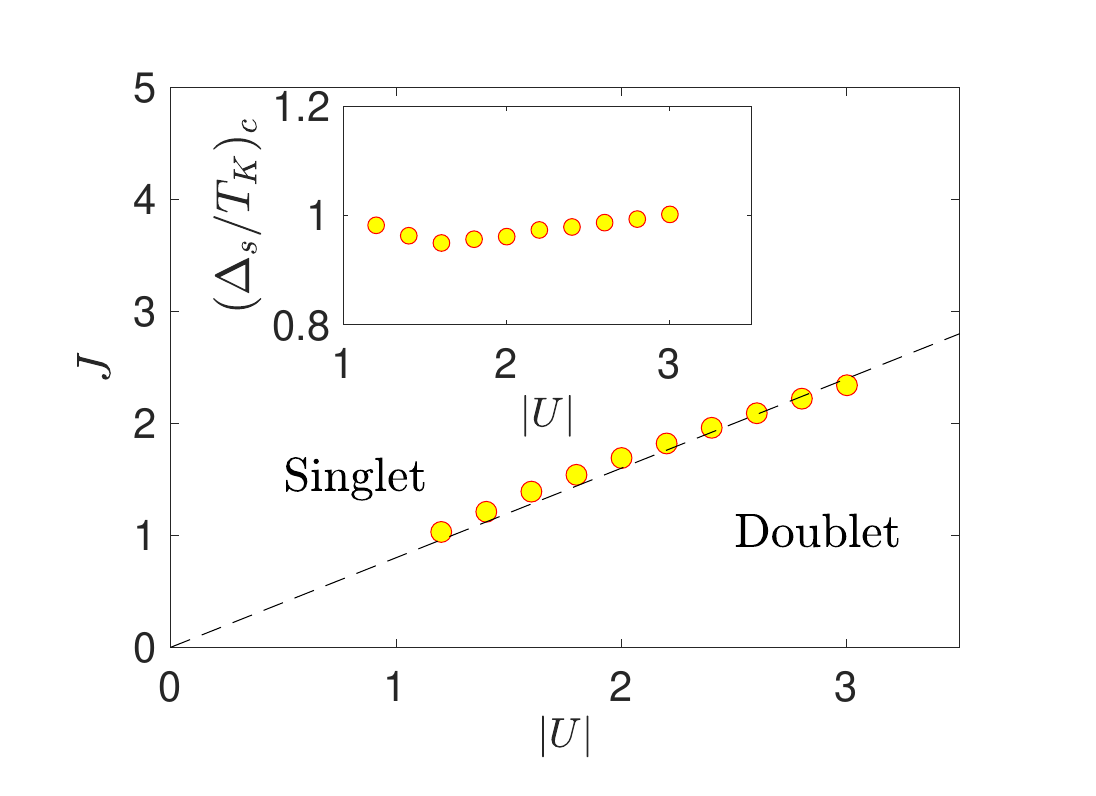}
    \caption{ The phase diagram of the model in the \((J, |U|)\) parameter space, obtained through DMRG calculations. The symbols 
    represent the critical couplings $J_c(U)$ at which the QPT occurs, as computed with the DMRG. 
     The dashed line serves as a visual guide, separating the many-body singlet ground state, where Kondo correlations dominate, from the doublet state, which features a free spin and residual superconducting correlations in the chain. The inset displays the critical
    ratio $(\Delta_s/T_K)_c$ calculated in the perturbative regime using Eqs.~\eqref{eq:Delta_s} and \eqref{eq:T_K}. The fitting parameter is fixed to $f=1.2$. The energy unit is the hopping amplitude $t$.}
    \label{fig:phase_diagram}
\end{figure}
For finite and negative \(U < 0\), superconducting correlations and Kondo screening coexist and compete in forming the ground state. The model is no longer integrable, necessitating numerical approaches to determine the ground state (GS). The preferred method is DMRG~\cite{white1992,schollwock2005} within the MPS formalism~\cite{schollwock2011}. We 
 solve the full Hamiltonian \eqref{eq:H} on chains with up to 400 sites using the ITensor package~\cite{itensor}. 
A sketch of the phase diagram is presented in  Fig.~\ref{fig:sketch}. Although the model does not exhibit long-range superconducting order, the competition between superconducting correlations, described by $\Delta_s$,  and Kondo screening, with $T_K$ as the associated energy scale, induces a QPT when \(\Delta_s/T_K \approx 1\), similar to the case of a magnetic impurity embedded in a regular s-wave superconductor. When \(\Delta_s < T_K\), the Kondo coupling is strong enough to form a Kondo singlet 
$|0 \rangle_s$ as the ground state (in this notation index $s$ stands for the spin sector), in which the impurity spin is 
completely dissolved into the environment, $\langle {\bf S}\rangle = 0$. In the opposite limit, \(\Delta_s > T_K\), the residual 
superconducting correlations dominate, preventing the formation of the Kondo cloud, resulting in a ground state that is a doublet $|\sigma\rangle_s$ with 
an asymptotically free impurity spin ($\langle S_z\rangle = \pm 1/2$) decoupled from the chain. 

The \(U\)-dependence of the critical coupling \(J_c(U)\) at which the QPT occurs is shown in 
Fig.~\ref{fig:phase_diagram}. The dashed line serves as a visual guide, separating the singlet 
and doublet regimes. The inset displays the ratio \(\Delta_s/T_K\), calculated using 
Eqs.~\eqref{eq:Delta_s} and \eqref{eq:T_K}, indicating a transition at \(\Delta_s/T_K \approx 1\) with a fitting parameter set at \(f = 1.2\). 

By bosonizing the exchange term for the impurity~\cite{nersesyan},
it couples only to the spin sector, which is gapped. It is important to realize that the open boundary bosonization~\cite{cazalillaboson} is satisfied by the requiring
$\phi_{c,s}(0)=\phi_{c,s}(L)=0$, where the $\partial_x\phi_{c,s}(x)$ describes the long wavelength charge and spin density operator, respectively.
From this and the bosonic representation of the conduction electron spin density\cite{giamarchi},
we obtain the spin operator interacting with the impurity as
\begin{subequations}
\begin{gather}
s^+_0=\frac{\kappa^+_\uparrow\kappa_\downarrow}{2\pi\alpha} \exp\left(-i\sqrt 2 \Theta_s(0)\right),\\
 s^z_0=\frac{\sqrt 2}{\pi}\partial_x\phi_s(x\rightarrow 0),
\end{gather}
\end{subequations}
which
indeed are independent from the charge fields, $\kappa_\sigma=\sum_{r=\pm}\kappa_{r,\sigma}$.

Therefore, the ensuing physics is reminiscent to that of a Kondo impurity coupled to a BCS superconductor, with the addition of a gapless charge mode,
decoupled from the impurity.
However, when physical observables are considered, both spin and charge degrees of freedom appear usually, making the Kondo physics of the
negative $U$ Hubbard model distinct from that
in a BCS superconductor as we show below. 
To sum up, we have a localized spin coupled to the gapped spin excitations described by Hamiltonian~\eqref{eq:H_s} and a gapless charge sector decoupled from 
the impurity described by Eq.~(\ref{eq:H_c}a).

\paragraph{Effective model.}
Similar to the Shiba problem~\cite{bauer2007}, close to the QPT, the system 
exhibits a subgap spectrum in the spin sector. Since our interest lies in the low-energy dynamics, we 
focus on the regime \(\omega \lesssim \Delta_s\). In this regime, it suffices to retain only the bound
 state arising from the interaction between the spin sector and the magnetic impurity.
The effective Hamiltonian in the spin sector is described in terms of the three states \(|0\rangle_s\) and \(|\sigma\rangle_s\) as 
\begin{equation}
%   H_s= \epsilon \sum_\sigma \big ( |\sigma\rangle_s \textmd{}_s\langle \sigma|-|0\rangle_s \textmd{}_s\langle 0|\big ),
H_{\rm Shiba}= \frac{\epsilon}{2}\big (|0\rangle_s \textmd{}_s\langle 0| - \sum_\sigma  |\sigma\rangle_s \textmd{}_s\langle \sigma|\big) ,
   \label{eq:H_s}
\end{equation}
where $\epsilon$ represents the energy splitting between the doublet and the singlet states. The singlet state 
$|0\rangle_s$ and the doublet states $|\sigma\rangle_s$ become  degenerate at the QPT.

 The effective Hamiltonian is therefore 
\begin{gather}
H_{\rm eff}=H_{\rm Shiba}+H_c.
\end{gather}
The total ground state wavefunction is written
for $J<J_c$ with $\epsilon<0$ using the  singlet state as $|GS\rangle=|0\rangle_s \otimes |N\rangle_c$, while 
 for $J>J_c$, $\epsilon>0$ and the total ground state wavefunction involves the doublet state with $|GS\rangle=|\sigma\rangle_s \otimes |N\rangle_c$.

\begin{figure}[t]
    \centering
    \includegraphics[width=0.9\columnwidth]{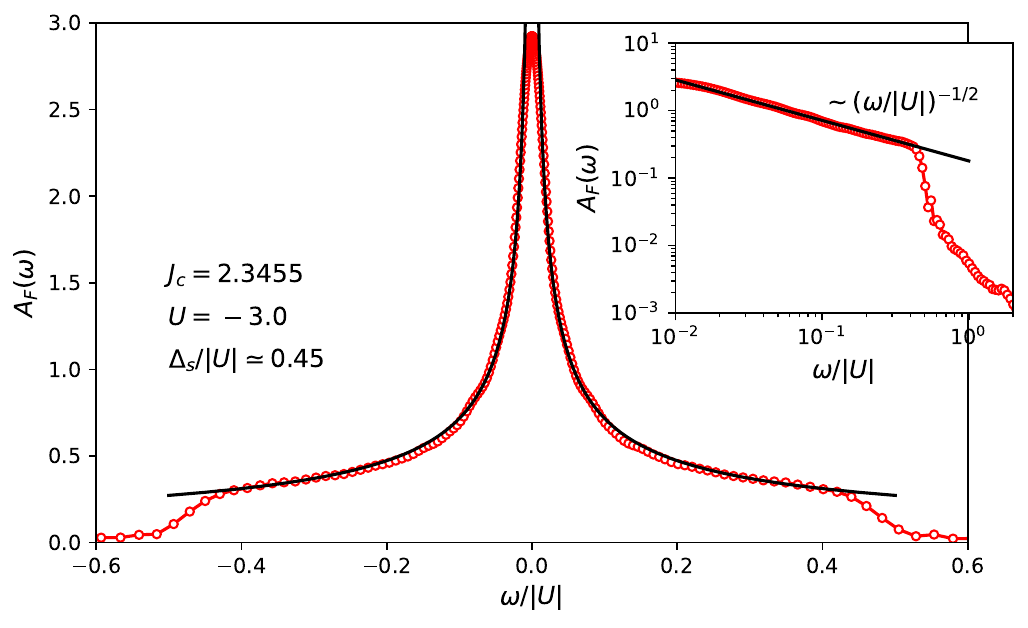}  
    \caption{ The spectral function $A_F(\omega)$ of the composite fermion operator at zero temperature, $(T=0)$,
      at the critical point, as function of frequency. 
      The symbols corresponds to the Chebyshev polynomial approach, while the solid 
    black line is the analytical result \eqref{eq:A_F} for $\epsilon=0$. 
     The Coulomb energy is fixed to $ U = -3.0 t$. The critical coupling as extracted from the DMRG ground state calculations is $ J_c \simeq 2.345 t $ and the spin gap is $\Delta_s\simeq 0.45|U|$. The calculation was performed using a chain of 100 sites. The inset shows the same data on a log-log scale, where the power-law behavior with exponent $-1/2$ is visible over almost two decades in energy.}.
    \label{fig:spectral_function}
\end{figure}

\paragraph{Composite fermion.} We investigate the spectral functions of
the composite fermion operator \cite{maltseva}, relevant for tunneling experiments~\cite{ruby2015tunneling,villas2021tunneling},
\begin{gather}
F^+_\sigma=\sum_{\sigma'}c^+_{0,\sigma'}{\bm \sigma}_{\sigma',\sigma}\cdot \bf S,
\end{gather}
giving $F^+_\uparrow = S^zc^+_{0,\uparrow}+ S^+ c^+_{0,\downarrow}$. Upon bosonization, it factorizes into charge and spin contributions as
$\sim \exp(-i\Theta_c(0)/\sqrt 2) ~ O^+_{s\uparrow}$, where $O^+_{s\uparrow}$ depends only on the spin degrees of freedom including the magnetic impurity as well as
\begin{gather}
O^+_{s\uparrow}=\frac{1}{\sqrt{2\pi\alpha}}\left(S^z\kappa^+_\uparrow\exp(-i\Theta_s(0)/\sqrt 2) \right.+\nonumber\\
\left. S^+\kappa^+_\downarrow\exp(i\Theta_s(0)/\sqrt 2)\right),
\end{gather}
and a similar expression for $O^+_{s\downarrow}$,
though its explicit form is not important for our calculations.
Its time dependent autocorrelator  is calculated for $J<J_c$ as
\begin{gather}
C_{F_\uparrow}(t)=\langle GS|F_\uparrow(t)F^+_\uparrow(0)|GS\rangle=\textmd{}_s\langle 0|O_{s\uparrow}(t)O^+_{s\downarrow}(0)|0\rangle_s\times\nonumber\\
\times_c\langle N|\exp(i\Theta_c(t,0)/\sqrt 2)\exp(-i\Theta_c(0,0)/\sqrt 2)|N\rangle_c=\nonumber\\
= \frac{\left|\textmd{}_s\langle \sigma|O_{s\uparrow}|0\rangle_s\right|^2}{2\pi\alpha} \exp(i\epsilon t)
\left(\frac{\alpha}{v_c t}\right)^{1/2K_c}.
\label{corrcomp}
\end{gather}
Here we made use of the fact that the composite fermion operator creates 
transitions between different parity spin sectors, and has vanishing matrix element between identical states,
namely $_s\langle 0| O_{s\uparrow}|0\rangle_s=0$.
After Fourier transformation\cite{giamarchi,delft,nersesyan}, its spectral function exhibits an $1/2K_c-1$ divergence in the low frequency regime as 
\begin{gather}
A_{F}(\omega) \sim \left(\omega-|\epsilon|\right)^{\frac{1}{2K_c}-1}
\label{eq:A_F}
\end{gather}
for $\Delta_s> \omega\geq |\epsilon|$. At half filling, $SU(2)$ symmetry dictates $K_c=1$, which gives the exponent $-1/2$, namely only half of the complex fermion contribution,
the other half from the spin sector gets gapped out and does not contribute for energies smaller than the spin gap.
Exactly at the critical point $J=J_c$, $\epsilon=0$ and the power law divergence starts at zero frequency.
This is coined as \emph{fractionalized Shiba state}. Typical behavior 
for the spectral function at the critical point is shown in Fig.~\ref{fig:spectral_function}.

Let us put this into perspective: for $J=U=0$, the impurity gets decoupled from the conduction electrons, which become non-interacting. Consequently,
the composite fermion autocorrelator in Eq. \eqref{corrcomp} becomes $t^{-1}$ and the ensuing spectral function is a frequency independent constant.
By reintroducing the interaction  and the Kondo coupling, the spin sector gets gapped out and a Shiba-type subgap state appears. The spin fractionalized part of the conduction electrons disappears from the composite fermion
correlator, and we are left with the $t^{-1/2}$ decay from the charge sector, pushed to the edge of the subgap state in the spin sector.
 Similar calculations apply in the doublet phase as well as for $F^+_\downarrow$.
We also point out that for $K_c<1/2$, the divergence at the excitation edge $|\epsilon|$ changes to a power-law vanishing spectral function. 

The key distinction between the \emph{regular} and \emph{fractionalized Shiba states} lies in the nature of the  gap and the behavior of the states within it. In the regular case, there is always a fixed, hard superconducting gap $\Delta$, with Shiba states forming as discrete bound states at various energies $0\le \epsilon<\Delta$ within the gap. In contrast, for the fractionalized Shiba state, the charge sector fills the gap spectrum down to $\epsilon$, the magnitude of the gap itself is $2\epsilon$, and no bound states develop inside the gap. Furthermore, at the critical point, $\epsilon=0$, the gap vanishes, and the spectrum becomes gapless.

\paragraph{Spectral function at $T=0$.} We compute the spectral function numerically using Chebyshev polynomials approach~\cite{weisse2006, holzner2011, wolf2014}. Initially, we construct the ground state $|GS\rangle $ of the model \eqref{eq:H} and compute its ground state energy $E_{GS}$ using DMRG calculations~\cite{schollwock2011}. 
The local spectral density of the composite fermion operator is given as
\begin{gather}
    A_F(\omega) =   \langle GS|F_\sigma \delta(\omega-E_{GS}-H) F^\dagger_\sigma|GS\rangle\\
    \nonumber
    +\langle GS|F^\dagger_\sigma \delta(\omega+E_{GS}-H) F_\sigma|GS\rangle. 
\end{gather}
Since the spectrum of the Hamiltonian~\eqref{eq:H} is bounded, we shift and rescale the Hamiltonian to adjust the bandwidth to the interval $[-1,1]$, which is the domain of the Chebyshev polynomials. Subsequently, we employ the Lanczos recursion method to generate a sequence of moments. 
 Using the recursive relations for the Chebyshev polynomials: $T_{n+1}(x) = 2x T_{n}(x)-T_{n-1}(x)$ the corresponding moments are computed as $\mu_n = \langle u |u_n\rangle$, where $|u \rangle = F^\dagger_\sigma |GS\rangle $ and $|u_{n+1}\rangle = 2 H | u_n\rangle -|u_{n-1}\rangle $, with $|u_0\rangle = |u\rangle$ and $|u_1\rangle = H|u\rangle$. In our analysis, we compute up to 20,000 moments. 
With these moments, we reconstruct the spectral function for the composite fermion as
 \begin{gather}
    A_F(\omega) = {1\over \pi\sqrt{1-\omega^2}}\Big (\mu_0 +2\sum_{n=1}^\infty \mu_n T_n(\omega)  \Big),
 \end{gather}
where the sum is approximated with a truncation of the order $M\simeq 20000$ while using a Jackson kernel $g_n$ to multiply to moments $\mu_n$ to damp the Gibbs oscillations~\cite{holzner2011}. 
It is worth mentioning that this approach allows us to compute the local density of states (at the impurity site for example) or any other spectral function of a fermionic or bosonic operator or for exploring the topology of the  spectral gap~\cite{lado2019}.
This numerical approach is better suited for the critical point, where the spectrum has no gap and allows for a direct comparison with analytical results. However, away from the critical point, the comparison becomes significantly more challenging. This is because it requires extremely large chains, and the presence of the gap necessitates the calculation of an exceptionally large number of moments.
The findings are illustrated in Fig.~\ref{fig:spectral_function}, featuring a comparison with the analytical result derived in Eq.~\eqref{eq:A_F} using the bosonization technique. The spectral function $A_F(\omega)$ is computed at the QPT, corresponding to $\epsilon = 0$, and exhibits a power-law decay exponent of $-1/2$ over nearly two decades in frequency, aligning perfectly with the analytical predictions. 

\paragraph{Finite temperature.}

\begin{figure}[t]
    \centering
    \includegraphics[width=0.49\columnwidth]{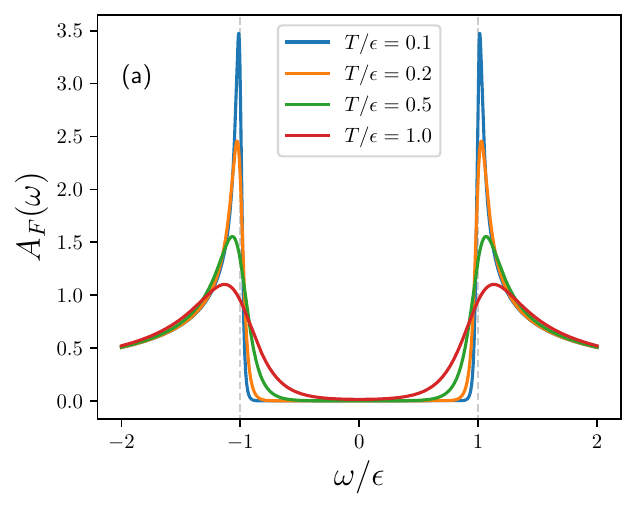}  
    \includegraphics[width=0.48\columnwidth]{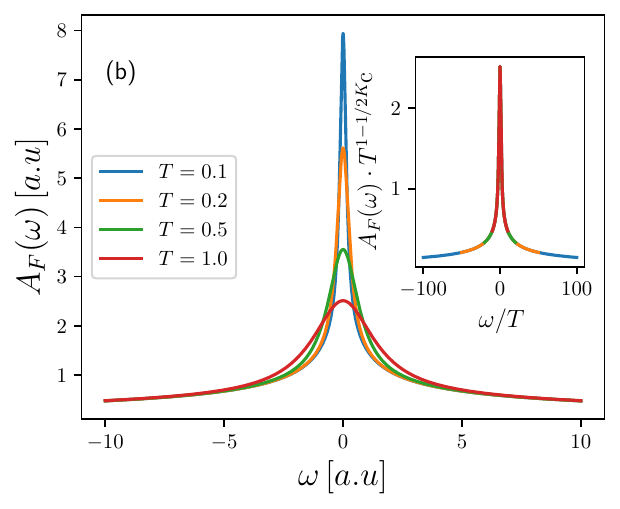}  
    \caption{ The finite temperature, ($T>0$), spectral function for the composite fermion as function of frequency away from the critical point (a) and   at the critical point (b) for different temperatures. 
    Inset: The universal function for the rescaled  spectral function \( A_F(\omega) \) of the composite fermion operator at finite temperatures at the critical point. For $\epsilon=0$ the only energy scale left in the problem is the temperature itself, and we use arbitrary units.}
    \label{fig:spectral}
\end{figure}
Finite temperature extension of Eq. \eqref{corrcomp} is also available by profiting from conformal invariance of the gapless 
charge sector~\cite{giamarchi,cazalillaboson}. 
At finite $T$, the density matrix is a direct product of the density matrices in the charge and spin sectors $\rho=\rho_c\otimes \rho_s$ with
\begin{equation}\label{eq:rho}
    \rho_s  = {1\over 1+2e ^{-\beta \epsilon}}\Big(\sum_{\sigma=\{\uparrow, \downarrow\}} e^{-\beta \epsilon} |\sigma\rangle_s \textmd{}_s\langle \sigma|  +|0\rangle _s\textmd{}_s\langle 0|\Big),
\end{equation} 
with $\beta=1/T$, the inverse temperature.
Straigtforward calculations give for the frequency dependence of the retarded Green's function of the composite fermion operator
\begin{gather}
    G^R_{F_\uparrow}(\omega) = N(\beta)
    \Big\{ e^{-i \pi/4K_c} \Big( e^{-\beta\epsilon} G(\omega+\epsilon)\textmd{}_s\langle \downarrow|O_{s\uparrow}O_{s\uparrow}^\dagger |\downarrow\rangle_s \nonumber\\
     +G(\omega-\epsilon)\textmd{}_s\langle \uparrow|O_{s\uparrow}O_{s\uparrow}^\dagger |\uparrow\rangle_s
     \Big )\nonumber\\
     +e^{i \pi/4K_c} \Big( e^{-\beta\epsilon} G(\omega-\epsilon)\textmd{}_s\langle \downarrow|O_{s\uparrow}O_{s\uparrow}^\dagger |\downarrow\rangle_s \nonumber\\
     +G(\omega-\epsilon)\textmd{}_s\langle \uparrow|O_{s\uparrow}O_{s\uparrow}^\dagger |\uparrow\rangle_s
     \Big )\Big\},
\end{gather}
with the prefactor $N(\beta) = -i {\beta \over \pi^2 \alpha}{1\over 1+2e^{-\beta\epsilon}}\Big({2 \pi \alpha \over \beta}\Big)^{1/2K_c}$ and  $G(\omega) = B (-i{\beta\omega\over 2 \pi}+{1\over 4K_c}, 1-{1\over 2K_c})$,  $B(x,y) =\Gamma(x)\Gamma(y)/\Gamma(x+y) $. The spectral function is extracted as the imaginary part, $A_{F}(\omega) = -2Im G^R_{F_\uparrow}(\omega)$.
At the critical point $\epsilon=0$, the retarded Green's function reduces to 
$G^R_{F_\uparrow}(\omega) \propto -i  B (-i{\beta\omega\over 2 \pi}+{1\over 4K_c}, 1-{1\over 2K_c})$. 
Typical results for the spectral function at finite temperatures away from the critical point are displayed in Fig~\ref{fig:spectral}(a) where the clear
opening of the gap $\Delta_s\approx 2\epsilon$ is visible, while at the critical point (Fig.~\ref{fig:spectral}(b)) the gap is closed. At the QCP, we are left with the temperature as the only energy scale, and therefore we use arbitrary units in Fig.~\ref{fig:spectral}(b). 
When properly rescaled at the QCP, the spectral function it shows a universal behavior as function of $\omega/T$, 
\begin{gather}
    A_F(\omega) \propto T^{1/2K_c-1}f\Big({\omega\over T}\Big)
\end{gather}
with $f(x)$ a universal function displayed in the inset of Fig~\ref{fig:spectral}(b).
\paragraph{Conclusions.}%

% Our study sheds light on the rich interplay between superconducting correlations and Kondo physics in one-dimensional systems. Through a detailed exploration of the phase diagram, we have identified a quantum phase transition where the dominance between Kondo screening and superconducting correlations changes. This shift alters the ground state from a Kondo many-body singlet to a doublet state characterized by a decoupled impurity spin with persistent residual superconducting correlations. 
% We employed theoretical methods, including density matrix renormalization group and bosonization, to understand the distinct behaviors of charge and spin sectors influenced by the impurity. We have computed the spectral function of some representative operator for tunneling experiments and analyze ground state properties of the model. Under the influence of a finite negative $U$ the spin sector gets gapped, resulting in a Shiba-like subgap spectrum, thus disappearing from the composite fermion spectral function. Meanwhile  the gapless charge sector exhibits a $\omega^{-1/2}$ decay at half-filling characteristic of a non-interacting chain. Extension to finite temperature indicates that at the quantum critical point the spectral function displays a universal behavior as function of $\omega/T$.
% Our findings are experimentally testable and may provide a pathway for investigating the fractionalized Shiba state. 

We have explored the emergence and properties of fractionalized Shiba states in a one-dimensional superconductor with fractionalized excitations. 
Unlike traditional Shiba states in bulk superconductors, the fractionalized Shiba states reflect the interplay between gapless charge and gapped 
spin excitations, characteristic of one-dimensional systems. Our analysis demonstrates that a side-coupled magnetic impurity interacts exclusively
 with spin degrees of freedom and induces a quantum phase transition despite the charge sector remaining decoupled. 

Through analytics and numerics, such as open boundary bosonization and using DMRG with the Chebyshev polynomial method, we have characterized the spectral properties of the fractionalized Shiba states. 
At zero temperature, the tunneling spectrum exhibits universal power-law scaling with an exponent of $-1/2$ at half filling, driven by the gapless charge excitations forming a Luttinger liquid. 
Finite-temperature extensions reveal that these spectral features maintain universal scaling behavior at the quantum critical point.

Our results provide an understanding of how magnetic impurities interact with fractionalized excitations in one-dimensional superconductors, giving rise to unique phenomena such as continuous 
midgap singularities. These findings have broader implications for the study of quantum phase transitions and open pathways for future research on interacting topological phases and possibly quantum-computational platforms based on coupled Shiba states.

\begin{acknowledgments}
This research is supported by the National Research, Development and
Innovation Office - NKFIH  within the Quantum Technology National Excellence
Program (Project No.~2017-1.2.1-NKP-2017-00001), K134437, K142179 by the BME-Nanotechnology
FIKP grant (BME FIKP-NAT).  C.P.M acknowledges support from CNCS/CCCDI–UEFISCDI, under projects number PN-IV-P1-PCE-2023-0159 and PN-IV-P1PCE-2023-0987.
\end{acknowledgments}

\bibliography{wboson1}

\end{document}